# Real-space observation of vibrational strong coupling between propagating phonon polaritons and organic molecules

Andrei Bylinkin[1,2], Martin Schnell[1,3], Marta Autore[1], Francesco Calavalle[1], Peining Li[1,4], Javier Taboada-Gutièrrez[5,6], Song Liu[7], James H. Edgar[7], Fèlix Casanova[1,3], Luis E. Hueso[1,3], Pablo Alonso-Gonzalez[5,6], Alexey Y. Nikitin*[2,3], and Rainer Hillenbrand*[8,3]

1 CIC nanoGUNE BRTA, 20018 Donostia - San Sebastian, Spain
2 Donostia International Physics Center (DIPC), 20018 Donostia-San Sebastián, Spain
3 IKERBASQUE, Basque Foundation for Science, 48009 Bilbao, Spain
4 Wuhan National Laboratory for Optoelectronics & School of Optical and Electronic Information, Huazhong University of Science and Technology, Wuhan 430074, China.
5 Departamento de Fisica, Universidad de Oviedo, 33006 Oviedo, Spain
6 Nanomaterials and Nanotechnology Research Center (CINN), El Entrego, Spain
7 Tim Taylor Department of Chemical Engineering, Kansas State University Manhattan, KS 66506, USA
8 CIC nanoGUNE BRTA and Department of Electricity and Electronics, UPV/ EHU, 20018 Donostia-San Sebastián, Spain

Corresponding author: * r.hillenbrand@nanogune.eu, alexey@dipc.org

**Phonon polaritons (PPs) in van der Waals (vdW) materials can strongly enhance light-matter interactions at mid-infrared frequencies, owing to their extreme field confinement and long lifetimes[1–7]. PPs thus bear potential for vibrational strong coupling (VSC) with molecules. Although the onset of VSC was observed spectroscopically with PP nanoresonators[8], no experiments have resolved VSC in real space and with propagating modes. Here, we demonstrate by nanoimaging that VSC can be achieved between propagating PPs in thin vdW crystals (h-BN) and molecular vibrations in adjacent thin molecular layers. We performed near-field polariton interferometry, showing that VSC leads to the formation of a propagating hybrid mode with a pronounced anti-crossing region in its dispersion, in which propagation with negative group velocity is found. Numerical calculations predict VSC for nanometer-thin molecular layers and PPs in few-layer vdW materials, which could make propagating PPs a promising platform for ultra-sensitive on-chip spectroscopy and strong coupling experiments.**

Phonon polaritons (PPs) – light coupled to lattice vibrations – in van der Waals (vdW) crystals open up new possibilities for infrared nanophotonics, owing to their strong infrared field confinement, picosecond-long lifetimes[1–7] and tunability via thickness and dielectric environment[9–12]. Since PPs in many vdW materials spectrally coincide with molecular vibrational resonances, which abound the mid-infrared spectral range, PP are thus promising candidates for achieving vibrational strong coupling (VSC) for developing ultrasensitive infrared spectroscopy and modification of chemical reactions by altering the vibrational energy of molecules[13–17]. Indeed, analogously to molecular vibrational infrared spectroscopy employing plasmons in graphene nanoribbons[18], localized PPs in hexagonal boron nitride (h-BN) high-Q factor nanoresonators were recently coupled to molecular vibrations[8], allowing for ultra-sensitive far-field spectroscopy at the strong coupling limit. However, ultra-confined propagating PPs in unstructured layers have neither experimentally nor theoretically been explored for field-enhanced molecular vibrational spectroscopy. More generally, none of the ultra-confined propagating polaritons in a 2D material has been exploited so far experimentally for field-enhanced molecular vibrational spectroscopy, although theoretical studies predict intriguing on-chip spectroscopy applications[19]. Further, real-space nanoimaging of the hybrid modes has been elusive, although it is of fundamental importance for in-depth experimental analysis of VSC exploiting PPs.

Here, we perform mid-infrared nanoimaging experiments[3–7] as a test bench to study the interaction of ultra-confined propagating polaritons in vdW materials with molecular vibrations in sub-100 nm thick organic layers. Specifically, we perform phonon-polariton interferometry of PPs in thin continuous h-BN layers interacting with 4,4′-Bis(N-carbazolyl)-1,1′-biphenyl (CBP) molecules. In contrast to typical strong coupling experiments, such as far-field spectroscopy in Kretschmann-Raether configuration or of polariton nanoresonators, we monitor in real space the effect of molecular absorption on PPs, leading to dramatic modification of the PP propagation length and anomalous dispersion with negative group velocity. We retrieve - in good agreement - experimentally and theoretically the quasi-normal modes of the CBP-PP coupled system, revealing significant anti-crossing and mode splitting caused by strong coupling. A numerical study predicts that few-layer h-BN films may enable to reach strong coupling even in the case of atomically thin molecular layers, thus underlining the potential of PPs to become a platform for ultra-sensitive on-chip spectroscopy devices.

In Fig. 1a we illustrate the polariton interferometry experiment. We illuminate the metallic tip of a scattering-type scanning near-field optical microscope (s-SNOM) to launch PP modes in a thin hexagonal boron nitride (h-BN) layer, which is placed above a thin layer of CBP molecules. The tip-launched PPs propagate to the h-BN edge, reflect and propagate back to the tip. The resulting polariton interference is shown in Fig. 1b by a numerical simulation of the electric field distribution along the h-BN/CBP layer. Importantly, the PP field penetrates into the molecular layer allowing for significant interaction between PPs and molecular vibrations. By mapping and analyzing the polariton interference spectroscopically, we can study how the molecule-PP interaction modifies the polariton wavelength $\lambda_{PP}$ and propagation length $L_{PP}$.

In a first experiment, we recorded the tip-scattered field $E_{sca}$ (which is governed by the polariton interference) as a function of frequency $\omega$ (using the nano-FTIR spectroscopy technique, see Methods) and distance $x$ between tip and h-BN edge (Fig. 1d). We observe the typical signal oscillations (fringes) arising from the polariton interference, whose period $\lambda_{PP}/2$ decreases with frequency $\omega$[7]. Importantly, our data reveal interruptions of the fringes (significant reduction of the amplitude signals) at the frequencies of the molecular vibrational resonances, which are absent in the data obtained on h-BN layers without molecules (Fig. 1h), clearly indicating significant interaction between PPs and molecular vibrations.

To visualize the PP dispersion, we performed a Fourier transform (FT) of the polariton interference pattern along the x-axis, revealing two bright branches in the momentum-frequency, $q$-$\omega$, domain (Fig. 1e and i for h-BN/CBP and pure h-BN layer, respectively, see Supplementary Information SII for data processing details). As in previous experiments[6], both branches (labeled E and T) can be attributed to the fundamental PP slab mode (typically referred as to M0 mode) launched by either the h-BN edge or the tip, respectively (Supplementary Fig. S4). The E-branch directly reveals the momenta $q$ of the M0 mode, in contrast to the T-branch revealing $2q$, as the M0 mode propagates twice the distance x between edge and tip. We clearly see interruptions at the spectral position of the molecular vibrations, manifesting strong damping of the PPs by molecular absorption.

The significant interaction between molecular vibrations and PPs could lead to highly-sensitive and ultra-compact IR spectroscopy devices. To briefly discuss this interesting aspect, we isolate the near-field signal of the tip-launched mode, $s_{PP}(x,\omega)$, by filtering and directly reveal PP attenuation for each frequency $\omega$ (Fig. 1f, see Supplementary Information SII for data processing details). We observe that the PPs are less strongly excited and decay faster near the molecular vibrational resonances, demonstrating the possibility of detecting molecular vibrational signatures via the accumulated PP losses. The numerically calculated propagation lengths of the M0 mode in the presence and absence of the CBP layer (red curves in Fig. 1g,k, respectively) corroborate our experimental results qualitatively. We find a reduced propagation length at the CBP absorption bands (e.g. 0.5 µm at 1510 cm$^{-1}$) compared to about 2 µm on pure h-BN.

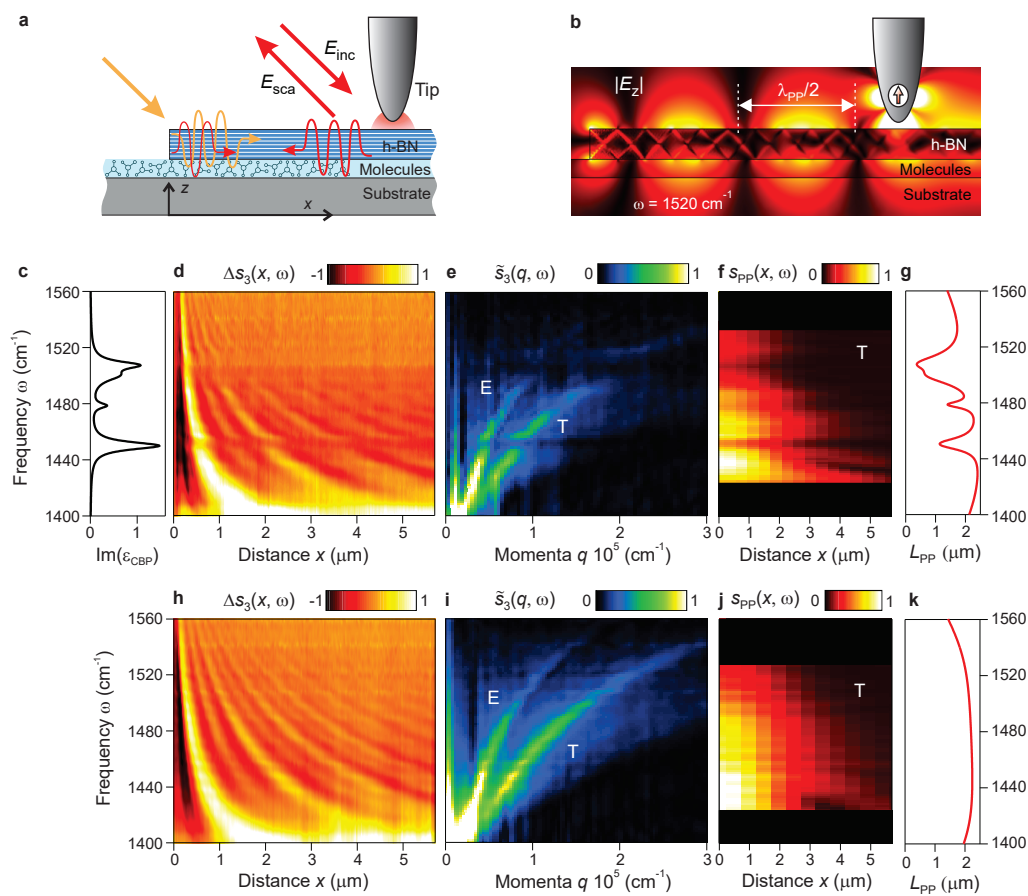

**Figure 1. Phonon-polariton interferometry of molecular vibrations a**, Illustration of the nanoimaging experiment. **b**, Simulated near-field distribution (z-component), generated by a point dipole source (red arrow) mimicking the illuminated tip. Outside the h-BN slab, it reveals essentially the interference of the fundamental PP slab mode (denoted M0), which is studied in this work. We note the appearance of a zig-zag pattern inside the h-BN slab. It is formed by superposition of higher order slab modes, which are typical for vdW polar crystals[1,2,6,10]. They propagate with significantly larger momenta, and thus do not contribute to the coupling between the M0 mode and the molecular vibrations studied in this work. **c,** Imaginary part

of the dielectric function $\varepsilon_{cbp}$ of CBP molecules **d, h**, Baseline-subtracted nano-FTIR amplitude signal $\Delta s_3(x,\omega)$ as a function of tip-edge distance $x$ for h-BN/CBP and pure h-BN layer. **e, i**, Amplitude of the Fourier transform of panel **d, h** along the $x$-axis. E and T mark the M0 mode excited by flake edge or tip, respectively. **f, j**, Isolated tip-scattered field of the M0 mode, obtained by inverse FT of the filtered T branches (Supplementary Information SII). **g, k** Theoretical propagation length $L_{PP}$ of the M0 mode. **b-g,** h-BN thickness is 50 nm, CBP thickness is 40 nm and substrate is 150 nm SiO$_2$ on Si. **h-k** h-BN thickness is 50 nm and substrate is 150 nm SiO$_2$ on Si.

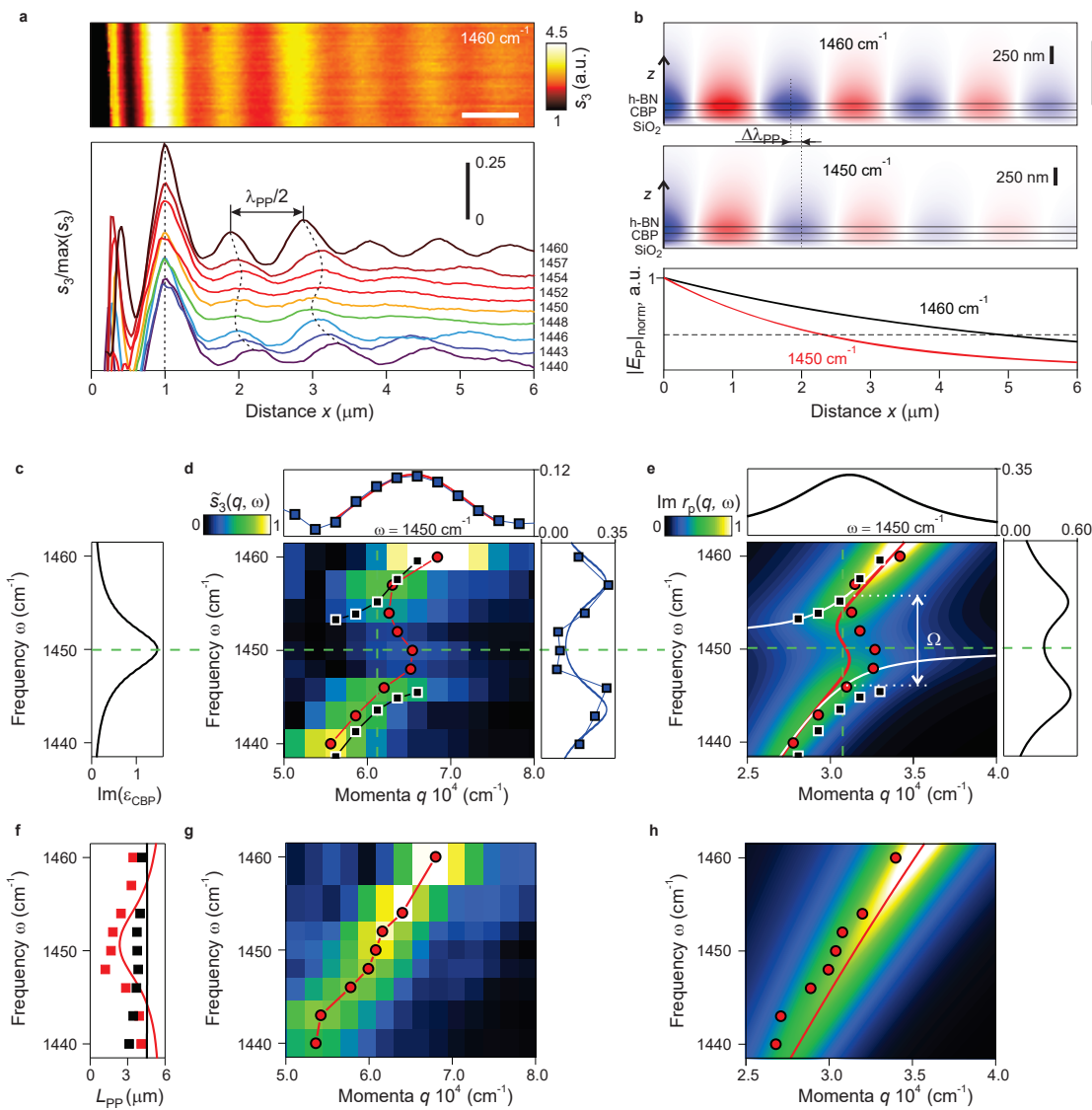

**Figure 2. Real space imaging of PPs on a h-BN/CBP layer in the region of anomalous dispersion. a,** (top panel) Infrared near-field amplitude image of a 85-nm-thick h-BN flake above a 100-nm-thick CBP layer at 1460 cm$^{-1}$. (bottom panel) Normalized amplitude profile perpendicular to the edge at different frequencies, extracted from images such as the one shown in the top panel. **b,** (top panels) Simulated near-field distribution (z-component) of PP mode propagating along the h-BN/CBP layer. (bottom panel) Calculated absolute value of the PP field as function of propagation distance $x$. **c,** Imaginary part of the dielectric function $\varepsilon_{CBP}$ of CBP molecules. **d,** Color-plot shows the amplitude of the FT of the line profiles of Fig. 2a. **e,** Color-plot $\tilde{s}_3(q,\omega)$ shows calculated imaginary part of Fresnel reflection coefficient. Curves

inside color plot show calculate dispersions assuming complex-valued momentum (red) or complex-valued frequency (white). White arrow indicates mode splitting Ω. **d,e,** Blue symbols and black curves in the right and top panels show line profiles along the dashed vertical and horizontal lines, respectively. **d,** Red curve in top panel shows a Lorentz function fit and blue curve in right panel shows a coupled oscillators fit. **f,** Experimental (symbols) and calculated (lines) propagation length $L_{PP}$ of PPs on h-BN/CBP (red) and pure h-BN (black) layer. **g,h,** Analogous to **d,e,** for h-BN layer without CBP.

To quantify the coupling between molecule vibrations and PPs, we performed a second experiment with improved signal-to-noise ratio and spectral resolution. To that end, we imaged PPs on a 85 nm thick h-BN layer above a 100 nm thick CBP layer at various frequencies around the CBP vibration at 1450 cm$^{-1}$ employing a quantum cascade laser (QCL) (Fig. 2a, top; see Methods). From the images we extracted line profiles (Fig. 2a, bottom; see Methods), which let us immediately recognize anomalous dispersion[20–22], near the CBP resonance. We clearly see that the PP wavelength $\lambda_{PP}$ increases as the molecular resonance is crossed from 1446 to 1454 cm$^{-1}$ (indicated by black dashed lines in Fig. 2a). To analyze the corresponding dispersion, we assembled FTs of the line profiles into a $\tilde{s}_3(q,\omega)$ plot (Fig. 2d, see Supplementary Information SIII for data processing details). By fitting the $\tilde{s}_3(q,\omega)$ in q-direction using Lorentz function and marking the maxima (red symbols in Fig. 2d), we clearly see that molecular absorption introduces a back bending in the PP dispersion, in stark contrast to the PP dispersion observed for h-BN without molecules (Fig. 2g, obtained analogously to Fig. 2d for pure h-BN layer). Further, the propagation length is significantly reduced by the presence of molecules (red symbols in Fig. 2f). Both observations represent a fundamental landmark feature of strong interaction between a propagating mode and a dipolar excitation, here, for the first time, observed in real space for ultra-confined phonon polaritons coupled to molecular vibrations (which may be named phovibrons). The near-field simulation shown in Fig. 2b confirms the reduction of both wavelength and propagation length when frequency decreases from 1460 to 1450 cm$^{-1}$.

The significant back bending of the dispersion indicates strong coupling between the PPs and the molecular vibrations of CBP, which in the following we corroborate by quasi-normal mode analysis. To this end, we extracted $\tilde{s}_3(\omega)$ line profiles from Fig. 2d for fixed momenta $q$ and fitted the data with a coupled-oscillator model[8,23,24] (see Supplementary Information SIV), yielding the dispersion of the quasi-normal modes, $\omega_\pm(q)$ (black symbols in Fig. 2d)[25]. The right panel of Fig. 2d shows, by way of an example, the experimental line profile (blue symbols) and coupled oscillator fit (blue line) for the q-value marked by the vertical dashed line in the $\tilde{s}_3(q,\omega)$-plot. We observe a clear anti-crossing behavior of the quasi-normal modes at the CBP resonance and a mode splitting of Ω = 11 cm$^{-1}$. Considering the uncoupled PP and CBP linewidths of $\Gamma_{PP} = 8 \text{ cm}^{-1}$ and $\Gamma_{CBP} = 6.4 \text{ cm}^{-1}$, respectively (Methods), we find that the strong coupling condition $C \equiv \frac{\Omega^2}{\left(\frac{\Gamma_{CBP}^2}{2} + \frac{\Gamma_{PP}^2}{2}\right)} = 2.3 > 1$ is well fulfilled[26].

To interpret the $\tilde{s}_3(q,\omega)$ data and to verify the extracted experimental dispersions, we performed a theoretical eigenmode analysis. To that end, we calculated the Fresnel reflection coefficient, $r_p$, of h-BN/CBP and pure h-BN layers on the SiO$_2$/Si substrate employing the transfer matrix (TM) method[27] (see Methods). We find that the Im($r_p$) plotted as a function of real $q$ and $\omega$ (Fig. 2e,h) describes well the experimental data $\tilde{s}_3(q,\omega)$ (Fig. 2d,g), particularly peak positions, linewidths and the saddle point at the CBP resonance, demonstrating that spatially Fourier transformed spectral polariton interferometry can be interpreted analogously to momentum- and frequency-resolved surface plasmon resonance spectroscopy employing, for example, the classical Kretschmann-Raether configuration[26,28–31].

To obtain the eigenmode dispersions, we determined the poles of $r_p$, assuming either complex momenta $q + i\kappa$ (corresponding to spatially decaying modes with propagation length $L_{PP} = 1/\kappa$) or complex frequencies $\omega - i\gamma$ (corresponding to temporally decaying modes with lifetime $\tau = 1/\gamma$). For spatially decaying modes in h-BN/CBP layers, we find a continuous dispersion exhibiting back bending (compare red curve in Fig. 2e with red curve in Fig. 2h showing results for pure h-BN layer) and reduced propagation lengths $L_{PP}$ around the CBP resonance (compare red curve in Fig. 2f with black curve showing $L_{PP}$ for pure h-BN layer). In contrast, analysis of temporally decaying modes yields quasi-normal modes featuring anti-crossing (white curves in Fig. 2e) and mode splitting of $\Omega = 10$ cm$^{-1}$, which according to C = 1.9 indicates strong coupling. The excellent agreement between experimental dispersions and propagation lengths (solid lines vs. symbols in Fig. 2e,f) clearly demonstrates the unique capability of polariton interferometry for comprehensive and quantitative analysis of the coupling between ultra-confined propagating polaritons and dipolar excitations, here specifically revealing strong coupling between propagating phonon polaritons and molecular vibrations.

We deepen the insights into strong coupling between molecular vibrations and PPs by an eigenmode analysis via TM calculations (analog to Fig. 2e) for various h-BN and CBP layer thicknesses. We first consider a variation of the CBP layer thickness $d_{CBP}$ for a fixed h-BN thickness of 35 nm (Fig. 3b-f). We find that the region of anomalous dispersion widens when $d_{CBP}$ is increased, which comes along with a reduction of the PP propagation length $L_{PP}$ (i.e. increase of PP damping) that reaches a finite minimum at the molecular resonance (Fig. 3c). Further, we find that with increasing $d_{CBP}$ the mode splitting $\Omega$ (the smallest vertical separation between the real frequency of the quasi-normal modes) and thus the coupling strength between PPs and molecular vibrations increases (Fig. 3e), as simply more of the PP fields lies inside the CBP layer (illustrated in Fig. 3a). Interestingly, the life times are rather large, about 2 ps, for all considered $d_{CBP}$ (red curves in Fig. 3f, obtained from the eigenmode analysis describe above), which is close to the molecular resonance lifetime of 1.7 ps and emphasizes that the coupled mode becomes more determined by the character of molecular absorption near the resonance. We note that the lifetime of the coupled mode being slightly longer than the lifetime of the bare molecular excitation has been also found by real-space observation of exciton polaritons[32]. In the region of anomalous dispersion PPs propagate with negative group velocity, $v_g = d\omega/dq < 0$ (Fig. 2d and Fig. 3b), yet with finite propagation length (red symbols in Fig. 2f and Fig. 3c). A simulation of the decay of narrowband PP pulses illustrates this effect (Fig. 3d). In the presence of a molecular layer, we find that the maximum of the pulse envelope after propagation of $x = 1.5$ μm appears at negative times $t$. Importantly, the negative group velocity yields negative values for the life time $\tau$ when calculated according to $\tau = L_{PP}/v_g$ (black curves in Fig. 3f), implying that this lifetime determination – typically applied in polariton interferometry[3,4,6,33,34] – cannot be used in the case of anomalous dispersion. Instead, eigenmode analysis in the picture of a temporally decaying mode is required (red curves in Fig. 3f).

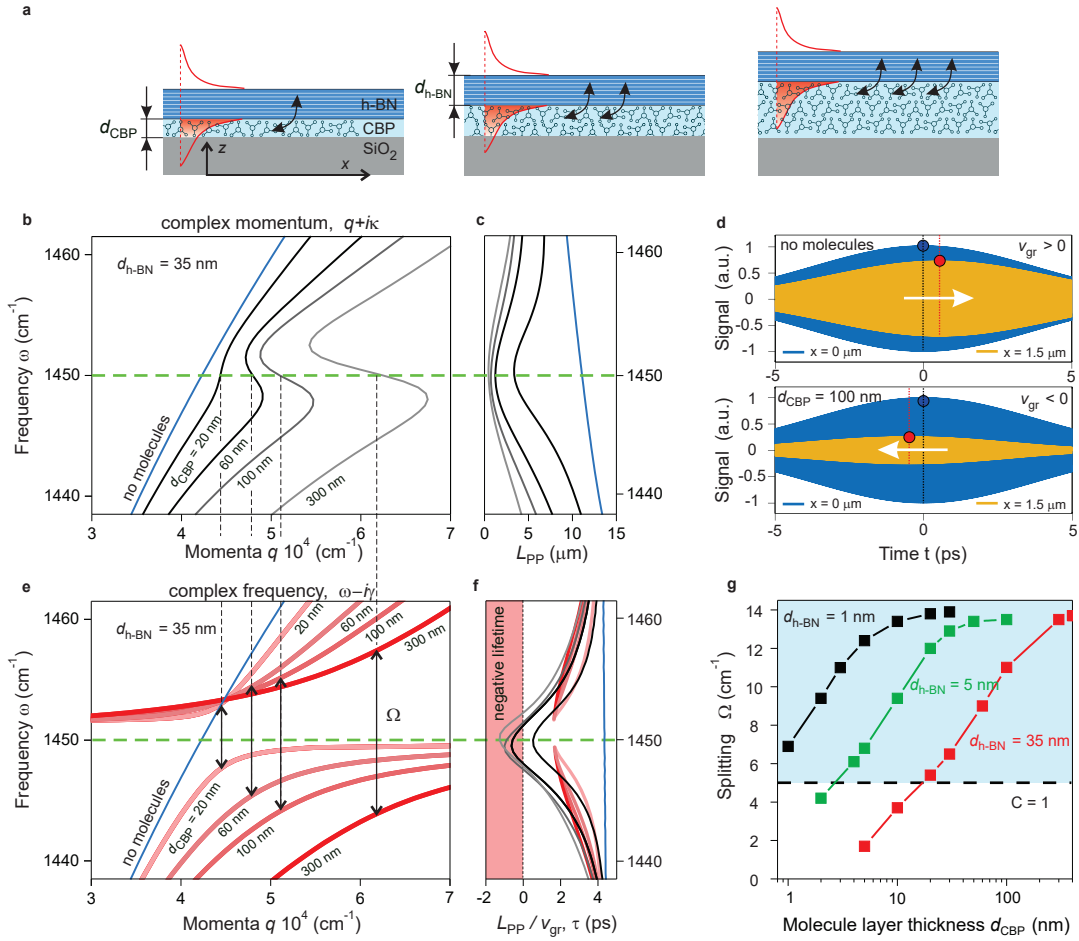

**Figure 3. Dispersion and mode splitting. a,** Illustration of phonon polaritons on differently thick molecular layers. **b,e,** Dispersions of PPs in a 35 nm-thick h-BN layer on CBP layers of various thicknesses $d_{CBP}$, obtained by eigenmode analysis via TM calculations assuming complex wavevectors $q + i\kappa$ (panel **b**) or complex frequencies $\omega - i\gamma$ (panel **e**). Horizontal green dashed line marks the frequency of the molecular vibrational resonance. **c,** PP propagation length. **d,** Narrowband PP pulses with 1450 cm$^{-1}$ center frequency and 3.5 cm$^{-1}$ spectral width at positions x = 0 and x = 1.5 µm in absence (top panel) and presence (bottom panel) of 100 nm-thick CBP layer below a 35 nm-thick h-BN layer. **f,** Blue and red lines show PP lifetime according to $\tau = 1/\gamma$. Black and gray lines show $L_{PP}/v_{gr}$. **g,** Mode splitting $\Omega$ as a function of the molecular layer thickness $d_{CBP}$ for h-BN layer with thickness $d_{h-BN}$ of 1 nm (black squares), 5 nm (green squares) and 35 nm (red squares). Horizontal black dashed line indicates the strong-coupling limit.

Plotting $\Omega$ as a function of $d_{CBP}$ (red symbols in Fig. 3g), we find saturation at about 14 cm$^{-1}$ for $d_{CBP} > 300$ nm, where the whole PP lies inside the CBP layer. Considering that the condition C > 1 in our simulations is fulfilled for $\Omega > 5$ cm$^{-1}$ (Supplementary Information SV, we find that strong coupling of CBP vibrations and PPs in a 35 nm tick h-BN layer can be already achieved for $d_{CBP}$ > 20 nm. Remarkably, further numerical mode analysis for h-BN layers with reduced thickness of $d_{h-BN} = 5$ nm and 1 nm (green and black symbols in Fig. 3g, respectively) predicts that in future experiments strong coupling may be achieved with just a few CBP monolayers ($d_{CBP} < 2$ nm). This can be explained by the extreme mode compression that comes along with the 100-fold reduced PP wavelength compared to the photon wavelength of the same energy[35].

Our work demonstrates that propagating PP in unstructured vdW materials can strongly couple to molecular vibrations, which could provide a platform for testing strong coupling and local controlling of chemical reactivity[13–16]. In addition, it opens up possibility for ultracompact, on-chip spectroscopy[19]. A large variety of vdW materials are at the disposal to go beyond the spectral range of h-BN, as recently demonstrated by PP observation on $MoO_3$[4] and $V_2O_5$[33]. In contrast to conventional polaritonic spectroscopy made with resonators (as, e.g. ribbons[8], cones[36] or hole arrays[37]) our work opens new avenues for studying strong-light matter interactions without the need of challenging sample structuring that typically comes along with additional losses from sample damage, inhomogeneities and scattering due to uncertainties in the sample fabrication process.

## Methods

### Sample preparation

4,4′-bis(N-carbazolyl)-1,1′-biphenyl with sublimed quality (99.9%) (Sigma-Aldrich, Saint Louis, MO, USA) was thermally evaporated in an ultra-high vacuum evaporator chamber (base pressure <$10^{-9}$ mbar), at a rate of 0.1 nm s$^{-1}$ using a Knudsen cell.

The h-BN crystal flake was grown from a metal flux at atmospheric pressure as described previously[38]. The thin layers used in this study were prepared by mechanical exfoliation with blue Nitto tape. Then we performed a second exfoliation of the h-BN flakes from the tape onto a transparent polydimethylsiloxane stamp. Using optical inspection of the h-BN flakes on the stamp, we identified high-quality flakes with appropriate thickness. These flakes were transferred onto a Si/SiO$_2$/CBP substrate using the deterministic dry transfer technique.

### nano-FTIR spectroscopy

We used a commercial scattering-type scanning near-field optical microscope (s-SNOM) setup equipped with a nano-FTIR module (Neaspec GmBH, Martinsried, Germany), in which the oscillating (at a frequency $\Omega \cong 270$ kHz) metal-coated (Pt/Ir) AFM tip (Arrow-NCPt-50, Nanoworld, Nano-World AG, Neuchâtel, Switzerland) is illuminated by p-polarized mid-IR broadband radiation generated by a supercontinuum laser (Femtofiber pro IR and SCIR; Toptica, Gräfelfing, Germany; average power of about 0.5mW; frequency range 1200–1700 cm$^{-1}$). ). The spectral resolution was set to 6.25 cm$^{-1}$, which is the limit of the setup. The spatial step (pixel) size in the nano-FTIR line scan shown in Fig. 1d is 20 nm , the total scan length is 6 μm and the number of pixels is 300. In Fig. 1g the pixel size is 27 nm , the total scan length is 8 μm and the number of pixels is 300. To suppress background scattering from the tip shaft and sample, the detector signal was demodulated at a frequency 3Ω. The nano-FTIR spectra were normalized to the spectra of the Si substrate, $s_3(x,\omega)e^{i\varphi_3(x,\omega)} = s_3^{h-BN/CBP}(x,\omega)e^{i\varphi_3^{h-BN/CBP}(x,\omega)} / s_3^{Si}(\omega)e^{i\varphi_3^{Si}(\omega)}$.

**Infrared nanoimaging by s-SNOM employing QCL illumination**

The oscillating tip (same parameters as described in nano-FTIR section) is illuminated by p-polarized mid-IR wavelength-tunable quantum cascade laser (QCL). The backscattered light is collected with a pseudo-heterodyne interferometer. To suppress background scattering from the tip shaft and sample, the detector signal was demodulated at a frequency $3\Omega$

Fig. 2a (top panel) shows a representative near-field amplitude image, $s_3$, of a 80-nm-thick h-BN flake above a 100-nm-thick CBP layer at $\omega = 1460$ cm$^{-1}$. By imaging the same sample area at various frequencies and averaging them in direction parallel to the h-BN edge, we obtained the near-field line profiles shown in Fig. 2a (bottom panel).

**Eigenmode analysis**

We used the transfer matrix approach to calculate the polariton eigenmodes[27]. They can be found by determining the poles in the Fresnel reflectivity of the layered sample for p-polarized light, $r_p$. To determine the poles, we numerically solved the equation $1/\text{Abs}(r_p) = 0$. For spatially decaying modes, we consider complex momenta $q + i\kappa$ and real-valued $\omega$, and determine the poles of $r_p(q+i\kappa, \omega)$, yielding $\omega(q)$ and the propagation length $L_{pr} = 1/\kappa$. For temporally decaying modes, we consider complex frequencies $\omega - i\gamma$ and real-valued $q$, and determine the poles of $r_p(q, \omega-i\gamma)$, yielding $\omega(q)$, lifetime $\tau = 1/\gamma$, and the mode linewidth $\Gamma = \gamma/2$. The dielectric permittivity of h-BN, CBP, Si and SiO$_2$ were modeled as described in Supplementary Information SI.

**Electromagnetic simulations**

Full-wave numerical simulations using the finite-elements method in frequency domain (COMSOL) were performed to study the electric field distribution around the h-BN/CBP heterostructure on top of a Si/SiO$_2$ substrate. The dielectric permittivity of h-BN, CBP, Si and SiO$_2$ were modeled as described in the Supplementary Information SI.

**Data availability**

The data that support the plots within this paper and other findings of this study are available from the corresponding authors on reasonable request.

**Acknowledgements**

The authors thank R. Esteban and J. Aizpurua for the fruitful discussions. The authors acknowledge financial support from the Spanish Ministry of Science, Innovation and Universities

(national projects MAT2017-88358-C3, RTI2018-094830-B-100, RTI2018-094861-B-100, and the project MDM-2016-0618 of the Maria de Maeztu Units of Excellence Program), the Basque Government (grant No. IT1164-19) and the European Union's Horizon 2020 research and innovation programme under the Graphene Flagship (grant agreement numbers 785219 and 881603, GrapheneCore2 and GrapheneCore3). F. Calavelle acknowledges support from the European Union H2020 under the Marie Sklodowska-Curie Actions (766025-QuESTech). J.T.-G. acknowledges support through the Severo Ochoa Program from the Government of the Principality of Asturias (no. PA-18-PF-BP17-126). P.A.-G. acknowledges support from the European Research Council under starting grant no. 715496, 2DNANOPTICA. Further, support from the Materials Engineering and Processing program of the National Science Foundation, award number CMMI 1538127 for h-BN crystal growth is greatly appreciated.

**Author contributions**

R.H. and M.A. conceived the study with the help of A.B and A.Y.N. Sample fabrication was performed by A.B. and F. Calavelle, supervised by F. Casanova and L.E.H. A.B. performed the experiments, data analysis and simulations. M.A., M.S. and J.T.-G. contributed to the near-field imaging experiments. M.A. and M.S. participated in the data analysis. P.L. contributed to simulations. S.L. and J.H.E. grew the isotopically enriched boron nitride. R.H. and A.Y.N. supervised the work. R.H., M.S. and A.B. wrote the manuscript with the input of A.Y.N., P.A.-G., M.A. All authors contributed to scientific discussion and manuscript revisions.

**Corresponding authors**

Correspondence to Rainer Hillenbrand or Alexey Y. Nikitin.

**Competing interests**

R.H. is co-founder of Neaspec GmbH, a company producing scattering-type scanning near-field optical microscope systems, such as the one used in this study. The remaining authors declare no competing interests.

# Supplementary information for "Real-space observation of vibrational strong coupling between propagating phonon polaritons and organic molecules"

## Table of Contents

# I. Dielectric function of materials

## A. CBP dielectric function

We measured the transmission spectrum of a 100 nm thick 4,4′-bis(Ncarbazolyl)-1,1′-biphenyl (CBP) layer evaporated on top of a CaF$_2$ substrate. To extract the dielectric function of the CBP molecules we used the following formula for thin films[1]:

$$\frac{T}{T_0} = \frac{1}{\left|1 + \sigma(\omega) d_{\text{CBP}} \frac{Z_0}{n_{\text{sub}} + 1}\right|^2} \quad (S1)$$

where $d$ is the film thickness ($d_{\text{CBP}} = 100$ nm $<< \lambda \approx 7$ μm), $n_{\text{sub}}$ is the refractive index of the substrate ($n_{\text{sub}} = 1.37$ for CaF$_2$ in the considered range), $Z_0$ is the impedance of free space (377 Ω) and $\sigma(\omega)$ is the complex conductivity of the thin film. $\sigma(\omega)$ is related to the permittivity, according to $\varepsilon(\omega) = 1 + \frac{i}{\omega \varepsilon_0} \sigma(\omega)$.

We modeled the dielectric function of the CBP molecules by the Drude−Lorentz model assuming 4 oscillators to describe the molecular vibrations as follows:

$$\varepsilon_{\text{CBP}}(\omega) = \varepsilon_\infty + \sum_k \frac{\omega_{p,k}^2}{\omega_{0,k}^2 - \omega^2 - i\Gamma_{\text{CBP},k}\omega}, k = 1 - 4 \quad (S2)$$

where $\varepsilon_\infty$ is high-frequency dielectric constant, $\omega_{p,k}$, $\omega_{0,k}$ and $\Gamma_{\text{CBP},k}$ represent the plasma frequency, the resonance frequency and the linewidth of the $k$-th Lorentz oscillators, respectively. We assumed that $\varepsilon_\infty = 2.8$, which is consistent with ellipsometry measurements reported in the literature[2] and which allows for proper fitting. Fig. S1 shows the relative transmission spectrum of a 100 nm CBP layer on top of CaF$_2$ and fit of the Eq. S1. All fit parameters are shown in Table S1.

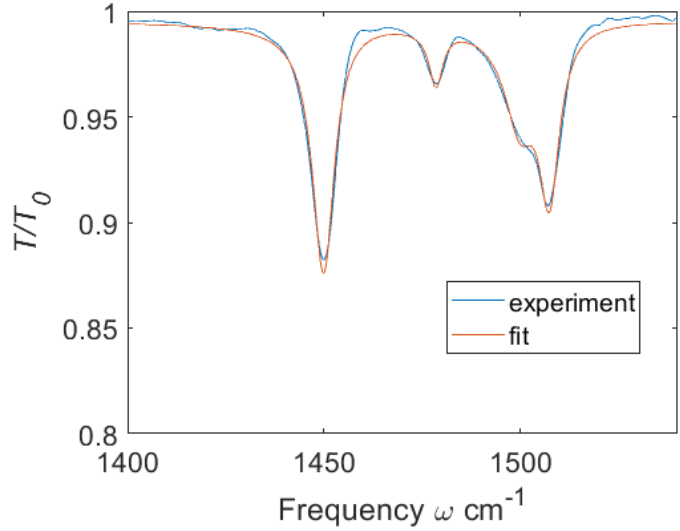

**Figure S1**. **Extraction of the dielectric function of the CBP molecules**. Blue line shows the relative transmission spectrum of the 100 nm layer of CBP molecules on top of the CaF$_2$ substrate. Red line shows the fit of Eq. S1.

| k | $\omega_{0,k}$ [cm$^{-1}$] | $\omega_{p,k}$ [cm$^{-1}$] | $\Gamma_{CBP,k}$ [cm$^{-1}$] |
|---|---|---|---|
| 1 | 1450 | 128 | 6.4 |
| 2 | 1478.6 | 47 | 4.4 |
| 3 | 1500.1 | 91 | 9.4 |
| 4 | 1507.4 | 99 | 6.1 |

**Table S1**. Fit parameters of Eq. S1 of the CBP dielectric function.

## B.    h-BN dielectric function

We used the isotopically ($^{10}$B) enriched h-BN[3]. The dielectric permittivity tensor of h-BN is modeled according to the following formula:

$$\varepsilon_{\text{h-BN},j}(\omega) = \varepsilon_{\infty,j}\left(\frac{\omega_{\text{LO},j}^2 - \omega^2 - i\omega\Gamma_j}{\omega_{\text{TO},j}^2 - \omega^2 - i\omega\Gamma_j}\right), \qquad (S3)$$

where $j = \perp, \parallel$ indicates the component of the tensor parallel and perpendicular to the anisotropy axis. We took the parameters for the dielectric function of h-BN from ref. 3 except of $\varepsilon_{\infty,\perp}$. We used $\varepsilon_{\infty,\perp} = 4.5$ instead of $\varepsilon_{\infty,\perp} = 5.1$ in ref. 3 for the best matching of our near-field experiments. We attribute the discrepancy to fabrication material parameter uncertainties. All parameters for the dielectric function, which were used in the simulation, are presented in the Table S2.

| j | $\varepsilon_\infty$ | $\omega_{TO}$ [cm$^{-1}$] | $\omega_{LO}$ [cm$^{-1}$] | $\Gamma$ [cm$^{-1}$] |
|---|---|---|---|---|
| $\perp$ | 4.5 | 1394.5 | 1650 | 1.8 |
| $\parallel$ | 2.5 | 785 | 845 | 1 |

**Table S2**. Parameters of the Eq. S3 of the dielectric function of h-BN.

## C.  Permittivity of the Si/SiO$_2$ substrate

We approximated the SiO$_2$ dielectric function taken data points from Palik[4], by following function:

$$\varepsilon_{SiO2} = p_1\omega^4 + p_2\omega^3 + p_3\omega^2 + p_4\omega + p_5, \quad (S4)$$

where $\omega$ is provided in units of wavenumbers. We took the following parameter: $p_1 = -2.27 \times 10^{-11} - 2.06 \times 10^{-12}i$, $p_2 = 1.50 \times 10^{-7} + 1.21 \times 10^{-8}i$, $p_3 = -3.74 \times 10^{-4} - 2.62 \times 10^{-5}i$, $p_4 = 0.42 + 0.025i$, $p_5 = -1.75 \times 10^2 - 8.816i$.

Simulations shown in Fig. 1 and 2 of the main text and in Fig. S4 consider the experimental Si/SiO$_2$ substrate, which is 150 nm-thick SiO$_2$ on highly doped Si ($\varepsilon_{Si++}$ = -11.11 + 10.49$i$). Note that the doped Si leads to the additional damping of PP. In future experiment the undoped Si should be used. Simulation shown in Fig. 3 of the main text consider a Si/SiO$_2$ substrate, which is 250 nm-thick SiO$_2$ on undoped Si ($\varepsilon_{Si}$ = 12).

## II. Data processing of nano-FTIR line scans

### A. Baseline subtraction from nano-FTIR line scans

We performed a baseline-subtraction from the nano-FTIR line scans in the $q$-$\omega$ domain to highlight the signal oscillations due to the polariton interference. Fig. S1a shows the amplitude of the nano-FTIR line scan, $s_3(x,\omega)$, of the 50 nm-thick h-BN flake with 40 nm-thick molecular layer below. We fitted the data set of Fig. 2d by a polynomial function of the $n = 1$ order in horizontal ($x$-axis) and $m = 2$ order in vertical ($y$-axis) directions, the fitted polynomial is:

$$\sum_{i=0}^{n}\sum_{k=0}^{m} a_{i,k} x^i y^k \tag{S5}$$

We obtained the following parameter for the polynomial fit: $a_{0,0} = 2.14 \times 10^{-6}$, $a_{1,0} = -4.04 \times 10^{-3}$, $a_{0,1} = -0.04$, $a_{1,1} = 133.03$, $a_{0,2} = 196.58$, $a_{1,2} = -993.93 \times 10^3$. Fig. S1b shows the subtracted polynomial function, $s_{\text{bg}}(x,\omega)$. Fig. S2c shows the baseline-subtracted nano-FTIR amplitude signal, $\Delta s_3(x,\omega)$, which is also shown in Fig. 1d of the main text.

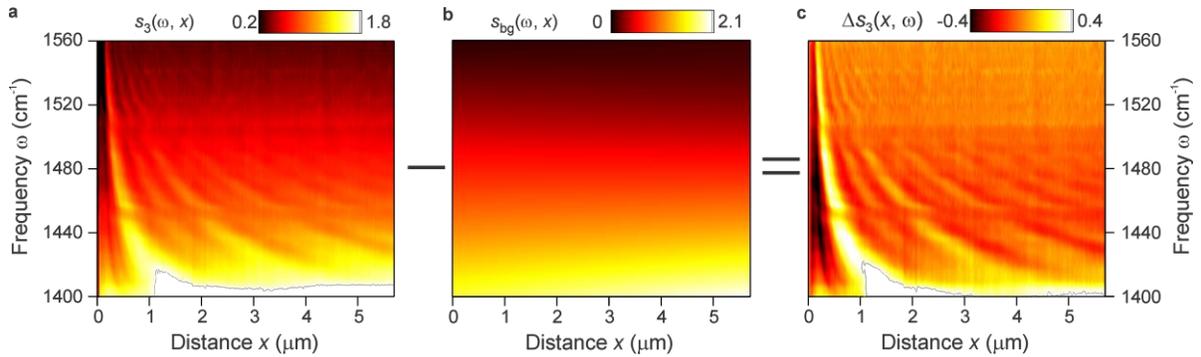

**Figure S2**. **Polynomial baseline-subtraction**. **a,** nano-FTIR amplitude signal $s_3(x,\omega)$ as a function of frequency and distance from the flake edge. **b**, Polynomial fit **c**, Baseline-subtracted nano-FTIR amplitude signal, $\Delta s_3(x,\omega)$.

We performed the same baseline subtraction procedure for the nano-FTIR line scan without molecules. We obtained the following parameter for the polynomial fit: $a_{0,0} = 2 \times 10^{-6}$, $a_{1,0} = -8.21 \times 10^{-3}$, $a_{0,1} = -28.3 \times 10^{-3}$, $a_{1,1} = 213.85$, $a_{0,2} = 102.56$, $a_{1,2} = -1.41 \times 10^6$. Baseline-subtracted nano-FTIR line scan amplitude is shown in Fig. 1h of the main text.
.

## B.   Fourier transform of the nano-FTIR line scans

Fig. S3a,b show the raw amplitude, $s_3(x,\omega)$, and phase, $\varphi_3(x,\omega)$, signals of the nano-FTIR line scan of the h-BN/CBP layers from Fig. 1 of the main text. The raw data comprise 300 pixels along the *x*-axis (distance *x* axis) and 156 pixels along the *y*-axis (frequency axis). For the Fourier transform (FT) of the nano-FTIR line scan along the *x*-axis, we prepared the nano-FTIR line scan data set according to the following procedure (see Fig. S3):

1. Remove the part of the data set, which was measured outside the flake (data on the left side of the white dashed line in Fig. S3a,b, which indicates the edge of h-BN layer). Fig. S3c shows the amplitude signal of data set after this step.
2. Join the data set from Fig. S3c with its mirror image along the *x*-axis. Fig. S3d shows the joint data set.
3. Perform baseline subtraction along the *x*-axis of the data set from Fig. S3d independently for each frequency. Fig. S3e shows the baseline-subtracted data set.
4. Apply window function (tukey window with cosine fraction 0.3, see Fig. S3f), $W$, to the data set from Fig. S3e. Fig. S3g shows the data set after windowing.
5. Zero-pad the data set from the Fig. S3g.

Fig. S3h shows the final data set after steps 1 to 5.

Note that in Fig. S3c-h we show only the amplitude of the nano-FTIR line scan, while data processing is done with the complex-valued data, that is, taking also into account the phase. The FT of the final data set yields Fig. 1e of the main text.

The raw complex-valued data set of the nano-FTIR line scan of h-BN without molecules was prepared in the same way before the FT. The FT of the final data set yields Fig. 1h of the main text.

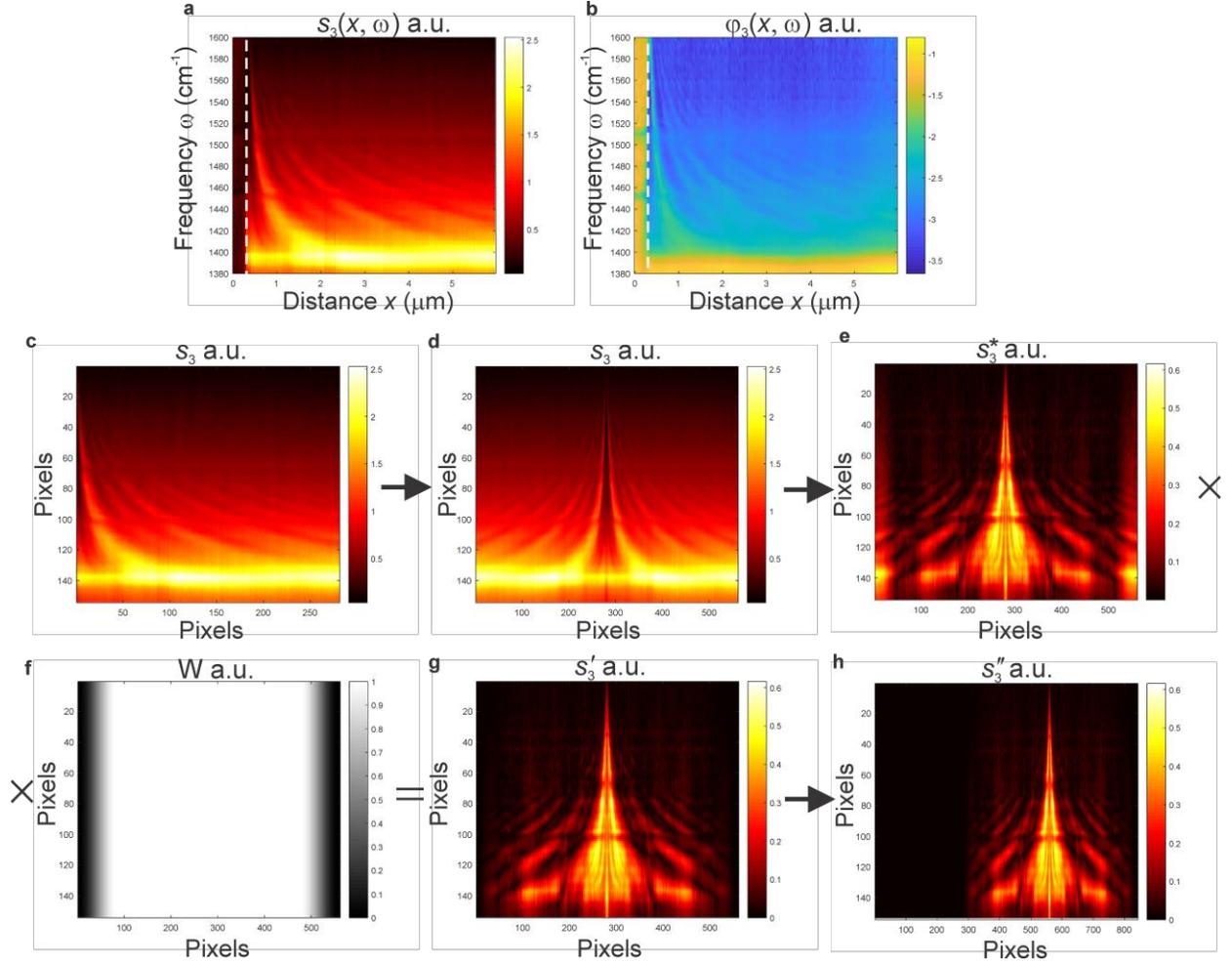

**Figure S3**. **Preparation of nano-FTIR line scan for FT**. **a,** Raw amplitude signal of the nano-FTIR line scan. **b,** Raw phase signal of the nano-FTIR line scan. **a,b**, Vertical dashed line indicates the h-BN flake edge. **c,** Amplitude signal of data set from only the h-BN/CBP layers structure. **d,** Amplitude signal of concatenated data set with mirrored data set. **e,** Baseline-subtracted amplitude signal of the data set. **f,** Window function. **g**, Amplitude signal of the data set after application of the window function. **h**, Final amplitude signal of the data set after zero padding.

### C.  Verification of the phonon polariton dispersion in Fig. 1 of the main text.

To describe the E- and T- branches seen in Fig. 1i,e of the main text, we calculated the PP dispersion of the fundamental PP slab mode (M0 mode), $q_{PP}(\omega)$, considering complex-valued momenta and real frequencies (see Methods). The E-branch corresponds to the M0 mode, which is excited by the edge of h-BN flake and propagates the distance between edge and tip once before it is scattered by the tip. As a result, the E-branch reveals the PP momenta and can be directly attributed to the PP dispersion, $q_E(\omega) = q_{PP}(\omega)$. The T-branch corresponds to the M0, mode which is excited by the tip and propagates the distance between edge and tip twice. As a result, the T-branch reveals the doubled PP momenta and can be attributed to the PP dispersion in the following way: $q_T(\omega) = 2 \times q_{PP}(\omega)$. Fig. S4 shows the comparison of the calculated PPs dispersion and $\tilde{s}_3(q, \omega)$.

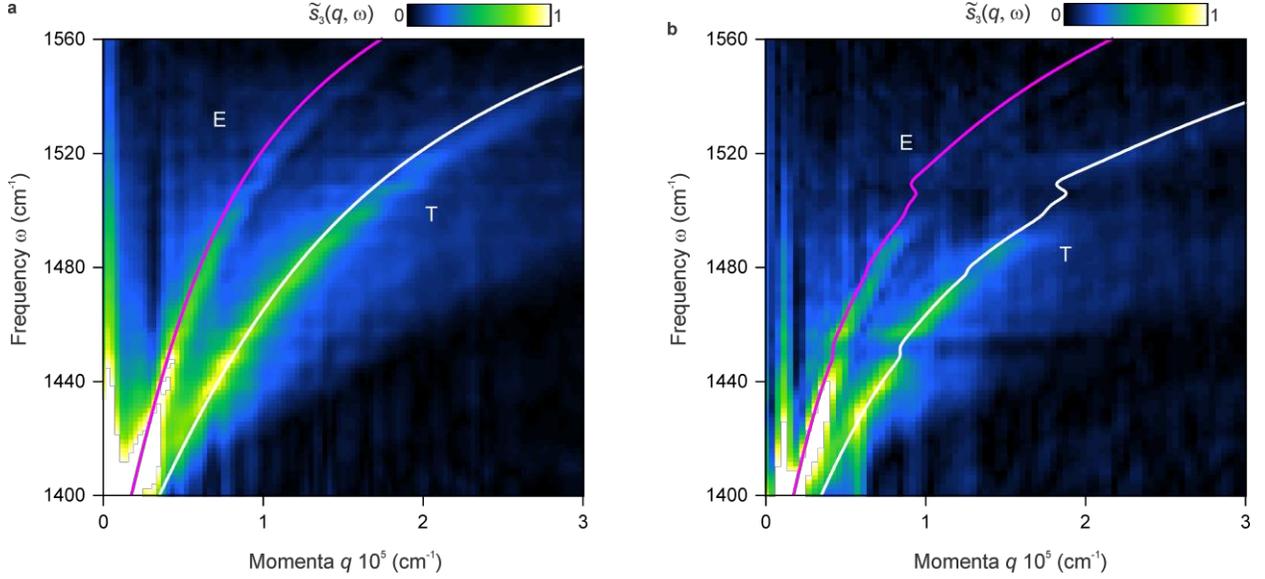

**Figure S4**. **Verification of the PP dispersions. a,b,** Colour-plots show the same data as the panels **i,e** of Fig. 1 of the main text, respectively. The magenta line shows the calculated PP dispersion, $q_E(\omega) = q_{PP}(\omega)$, which matches to the E-branch. The white line shows the calculated PP dispersion with doubled momenta, $q_T(\omega) = 2 \times q_{PP}(\omega)$, which matches to the T-branch. **a,** pure 50 nm-thick h-BN flake. **b,** 50 nm-thick h-BN with 40 nm-thick CBP layer below.

## D.    Inverse Fourier transform of the isolated M0 mode

To extract the tip-scattered PP field, we apply a filter in the $q$-$\omega$ domain (Fourier space) to isolate the tip-launched mode (T-branch in Fig. 1e,i of the main text), and perform an inverse Fourier transform (iFT) to obtain the tip-scattered field of the M0 mode in the $x$-$\omega$ domain[5]. This process is shown in Fig. S3, where the initial amplitude of the FT is shown in panel a, the applied filter function is shown in panel b, the filtered amplitude of the FT is shown in panel c, the result of the inverse FT is shown in panel d. The filter function $W(q,\omega)$ is a Gaussian function of the width $\Delta q = 0.4 \times 10^5$ cm$^{-1}$.

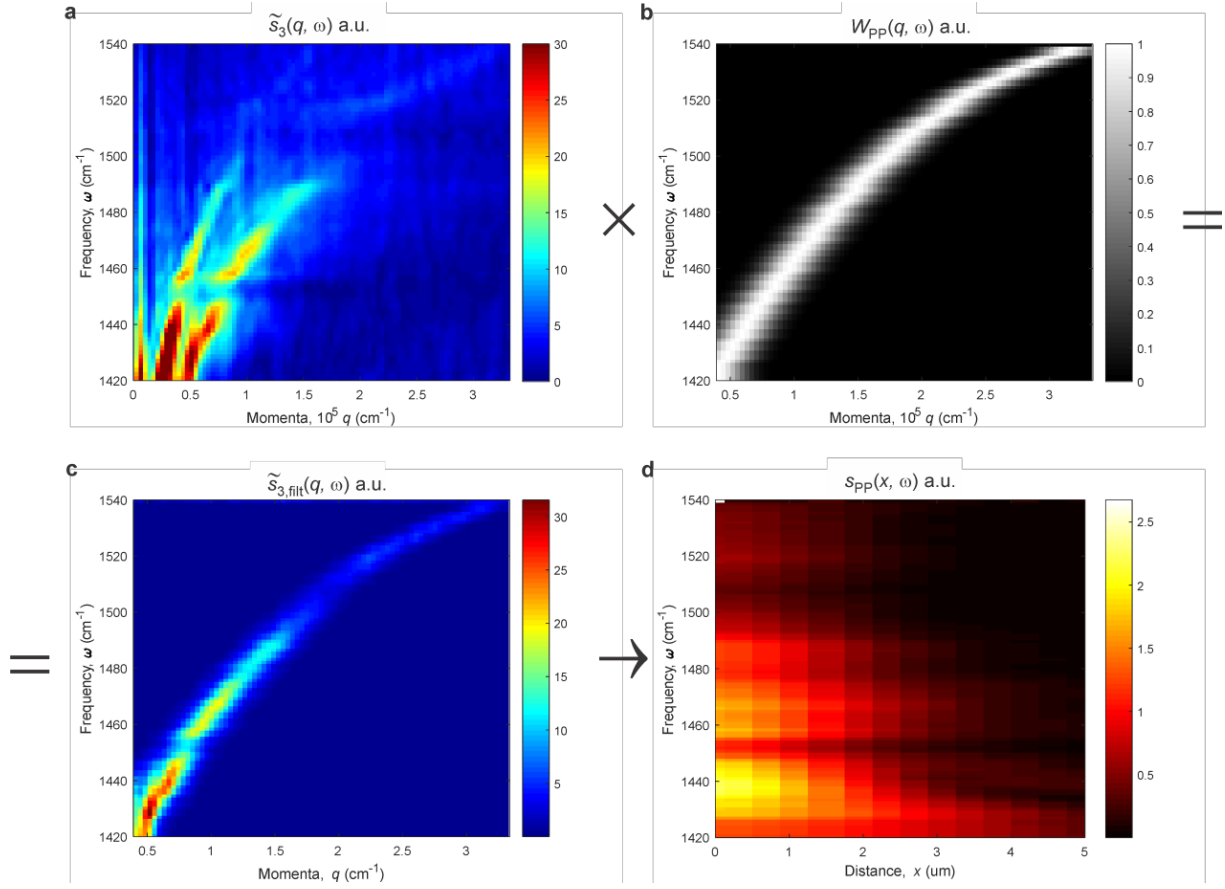

**Figure S5**. **Filtering the T-branch in the $q$-$\omega$ domain**. **a,** Amplitude of the FT in the $q$-$\omega$ domain. **b,** Filter function. **c,** Isolated T-branch after multiplication of panel **a** and panel **b**. **d**, Inverse FT of panel **c,** revealing the tip-scattered pure PP field as a function of $x$ and $\omega$.

## III.  Data processing of the line profiles from Fig. 2a

### A.  Fourier transform of the line profiles

We performed phonon polariton interferometry by s-SNOM employing a wavelength-tunable quantum cascade laser (QCL). By imaging the same sample area at various frequencies and averaging them in direction parallel to the h-BN edge, we obtained the line profiles for 85 nm-thick h-BN flakes with 100 nm thick layer of CBP molecules below (Fig. 2a of the main text) and pure 85 nm-thick h-BN without molecules.

To prepare our line profiles for the FT, we first normalized each amplitude line profile to the amplitude signal far away from edge. In second step, we cut the first oscillation in the amplitude line profile (first fringe) to avoid contributions of edge modes[6] to our FT analysis. The remaining steps of the data preparation process for the FT were the same to the data preparation of the nano-FTIR data set (see details in Supplementary information SII). Fig. 2d,g of the main text show the result of the FT of the line profiles of pure h-BN and h-BN/CBP layers, respectively.

## B.     Phase velocity of the propagating molecule-coupled phonon polariton mode

To check the sign of the PPs phase velocities in the region of anomalous dispersion ($\omega = 1450$ cm$^{-1}$) and compare with the phase velocities outside of the anomalous dispersion ($\omega = 1440$ cm$^{-1}$, 1460 cm$^{-1}$), we plot in Fig. S6 the s-SNOM line profiles in the complex plane. We observe an anticlockwise rotating spiral at all frequencies, revealing positive phase velocities in these regions, which is consistent with the previous result for the PPs[7].

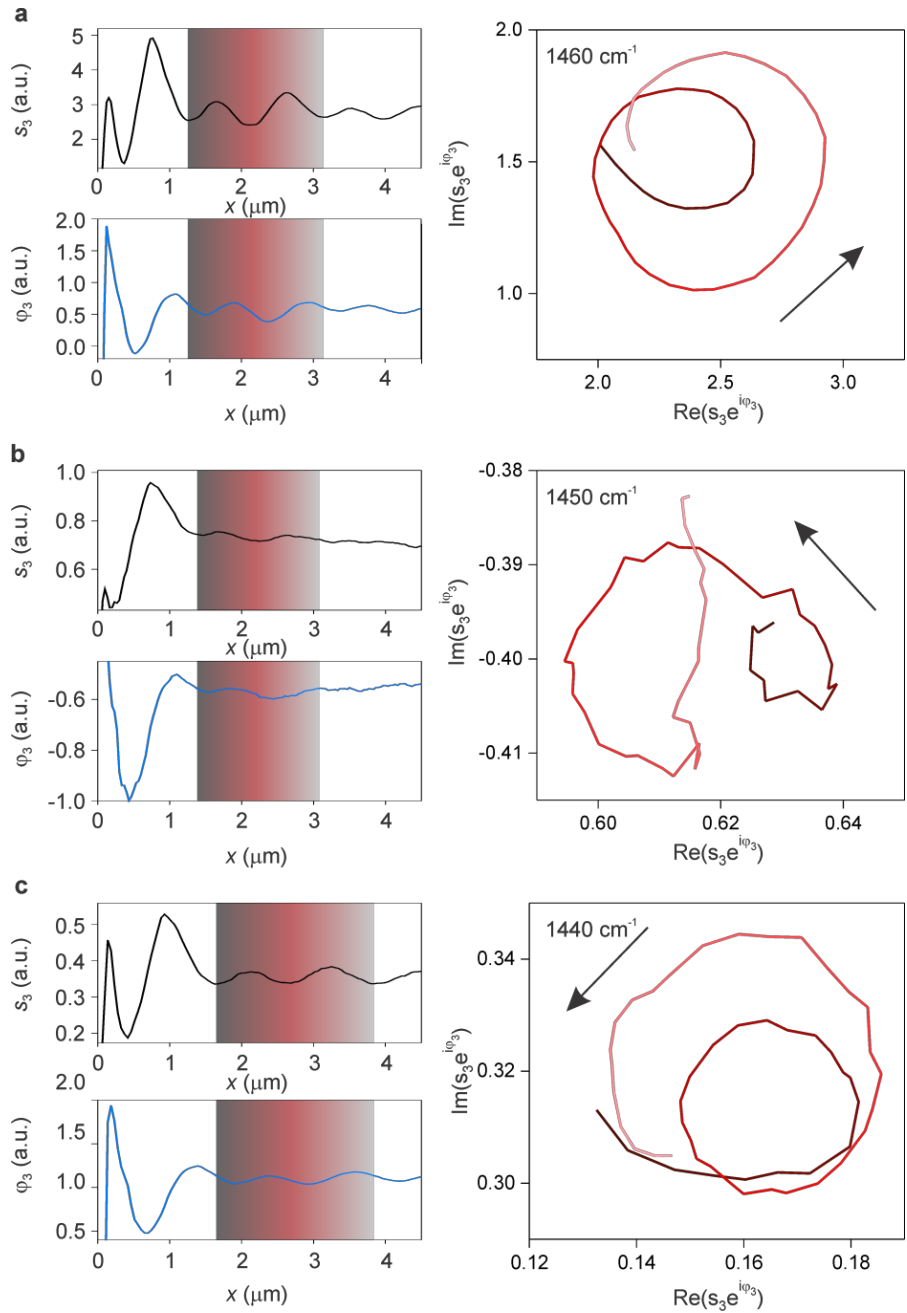

**Figure S6**. **s-SNOM line profiles plotted in the complex plane.** Black (blue) line shows the amplitude (phase) signal of the complex-valued s-SNOM signal as a function of the distance *x*. Gradient line (black-red-white) shows the signal in the complex plane corresponding the gradient area on the amplitude and phase line profiles. Black arrows indicate the anticlockwise direction of rotation, which corresponds to a positive phase velocity. **a,** $\omega = 1460$ cm$^{-1}$, **b,** $\omega = 1450$ cm$^{-1}$, **c,** $\omega = 1440$ cm$^{-1}$.

# IV. Classical coupled harmonic oscillator model

In order to analyze the vertical line profiles of $\tilde{s}_3(q,\omega)$ in Fig. 2d of the main text, we phenomenologically described the coupling of the molecular vibrations and the phonon polaritons via a classical model of coupled harmonic oscillators[8,9]. The equation of motion for the two coupled harmonic oscillators are given by[10]:

$$\begin{cases} \ddot{x}_{PP}(t) + \Gamma_{PP}\dot{x}_{PP}(t) + \omega_{PP}^2 x_{PP}(t) - 2g\bar{\omega}x_{CBP}(t) = F_{PP}(t) \\ \ddot{x}_{CBP}(t) + \Gamma_{CBP}\dot{x}_{CBP}(t) + \omega_{CBP}^2 x_{CBP}(t) - 2g\bar{\omega}x_{PP}(t) = F_{CBP}(t) \end{cases}$$

where $x_{PP}$, $\omega_{PP}$ and $\Gamma_{PP}$ represent the displacement, frequency and linewidth of the PP mode, respectively. $F_{PP}$ represents the effective force that drives its motion and is proportional to the external electromagnetic field. The corresponding notation is valid for the CBP vibration. $g$ represents the coupling strength and $\bar{\omega} = (\omega_{PP} + \omega_{CBP})/2$. The extinction, $C_{ext}$, of such system can be calculated according to $C_{ext} \propto \langle F_{PP}(t)\dot{x}_{PP}(t) + F_{CBP}(t)\dot{x}_{CBP}(t) \rangle$ [8].

We fit this model to the vertical line profiles of $\tilde{s}_3(q,\omega)$ (line profiles at fixed momentum). In the fitting procedure, $\Gamma_{CBP} = 6.5$ cm$^{-1}$ and $\Gamma_{PP} = 8$ cm$^{-1}$ were fixed according to the CBP dielectric function and eigenmode analysis (considering complex frequencies) of uncoupled PP in the h-BN, which is separated from Si/SiO$_2$ substrate by 100 nm spacer (SiO$_2$ thickness is 150 nm) (see methods and Supplementary info SV). $\omega_{CBP}$ was limited within a few wavenumbers from its initial value ($\omega_{CBP} = 1450$ cm$^{-1}$), to allow for an eventual Lamb shift of the molecular vibration[11,12]. $\omega_{PP}$ was considered as free parameters in all fits. The coupling strength $g$ was considered as a free parameter in three fits at $q = (5.86, 6.12, 6.36) \times 10^4$ cm$^{-1}$, which were shown by black curves in Fig. S7a. The extracted values of the coupling strength for each fits are plotted as black symbols in the Fig. S7b. Dashed black line in Fig. S7b indicates the average coupling strength of extracted values, $g_{avg} = 6.6$ cm$^{-1}$. We fixed the coupling strength to $g = g_{avg} = 6.6$ cm$^{-1}$ in the fits at $q = (5.62, 6.60) \times 10^4$ cm$^{-1}$, which are shown by red curves in Fig. S7a. With the parameters extracted from the fits we calculated the quasi-normal modes of the coupled system, according to[13]:

$$\omega_\pm + \frac{i\Gamma_\pm}{2} = \frac{\omega_{PP} + \omega_{CBP}}{2} - i\frac{\Gamma_{PP} + \Gamma_{CBP}}{4} \\ \pm \frac{1}{2}\sqrt{4g^2 + \left(\omega_{PP} - \omega_{CBP} - i\frac{\Gamma_{PP} - \Gamma_{CBP}}{2}\right)^2} \qquad (S6)$$

Fig. S7c shows the calculated quasi-normal modes $\omega_\pm$ as a function of $\omega_{PP}$. We clearly observe a mode splitting of $\Omega = 11$ cm$^{-1}$ at $\omega_{PP} = 1450$ cm$^{-1}$ (the smallest vertical separation between real frequencies of the branches). With $\Gamma_{CBP} = 6.5$ cm$^{-1}$ and $\Gamma_{PP} = 8$ cm$^{-1}$, the mathematical condition of strong coupling[14], $C \stackrel{\text{def}}{=} \frac{\Omega^2}{\frac{\Gamma_{CBP}^2}{2}+\frac{\Gamma_{PP}^2}{2}} = \frac{11^2}{\frac{6.4^2}{2}+\frac{8^2}{2}} = 2.3 > 1$, is fully fulfilled.

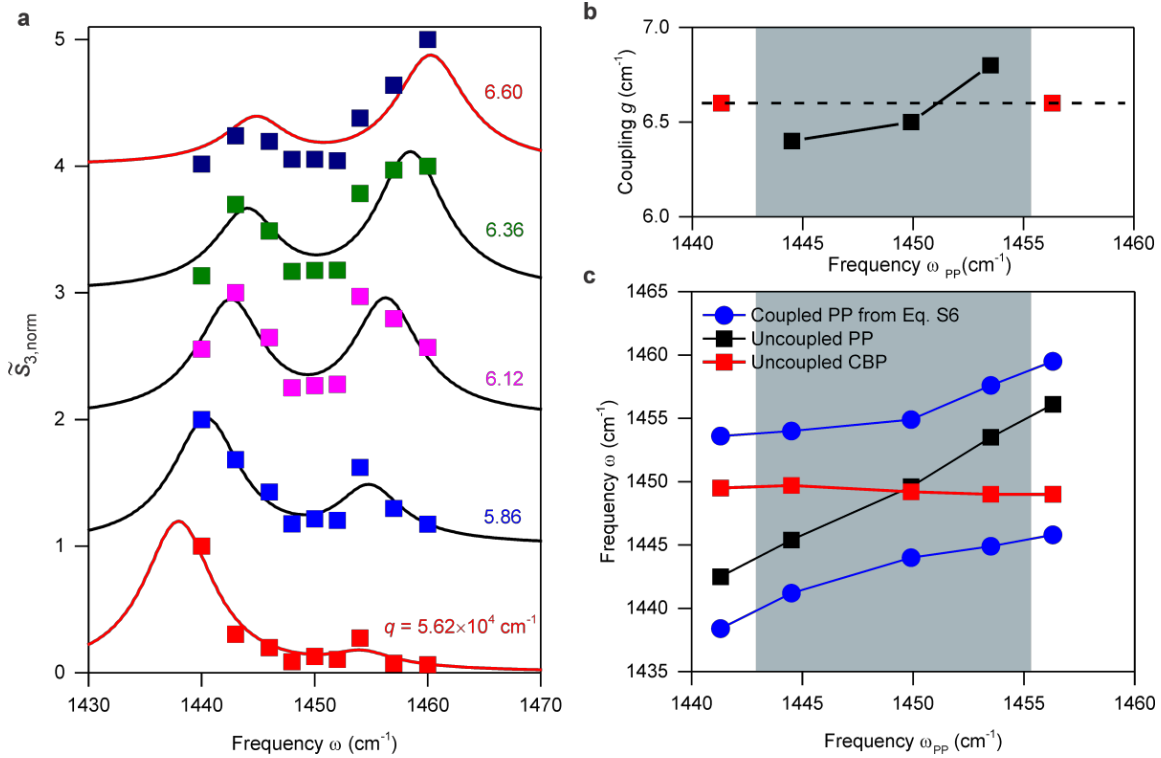

**Figure S7. Coupled-oscillator fit of vertical line profiles of the experimental data set of Fig. 2d of the main text,** Squares show normalized amplitude of FT as function of frequency for fixed momenta, $\tilde{s}_{3,\text{norm}}(\omega) = \tilde{s}_3(\omega)/\max(\tilde{s}_3(\omega))$. Data were extracted from Fig. 2d of the main text. Black lines are fit of the coupled oscillators model with coupling strength $g$ as a free parameter. Red lines are fit of the coupled oscillators model with fixed coupling strength $g_{\text{avg}} = 6.6$ cm$^{-1}$. **b,** Black symbols show the coupling strength from the fits, black dashed line indicates the average coupling strength, $g_{\text{avg}} = 6.6$. **c,** Frequencies of the quasi-normal modes of the coupled system calculated from Eq. S6, $\omega_{\pm}$, frequencies of the uncoupled PP, $\omega_{PP}$, and uncoupled CBP molecules, $\omega_{CBP}$, as extracted from the fit, plotted as a function of $\omega_{PP}$. **b,c,** Blue region indicates the points of the fits with coupling strength, $g$, as a free parameter.

# V. Theoretical strong-coupling analysis

To theoretically study the coupling between PPs and CBP molecular vibration we analyze systems illustrated in Fig. S8. We consider PPs in the pure h-BN slab with dielectric spacer below on top of the Si/SiO$_2$ substrate as uncoupled PP mode, polaritons in the h-BN/CBP layers as a coupled system. The CBP vibration at $\omega$ =1450 cm$^{-1}$ is modeled as Drude-Lorentz oscillator (see details in Supplementary information I, CBP dielectric function).

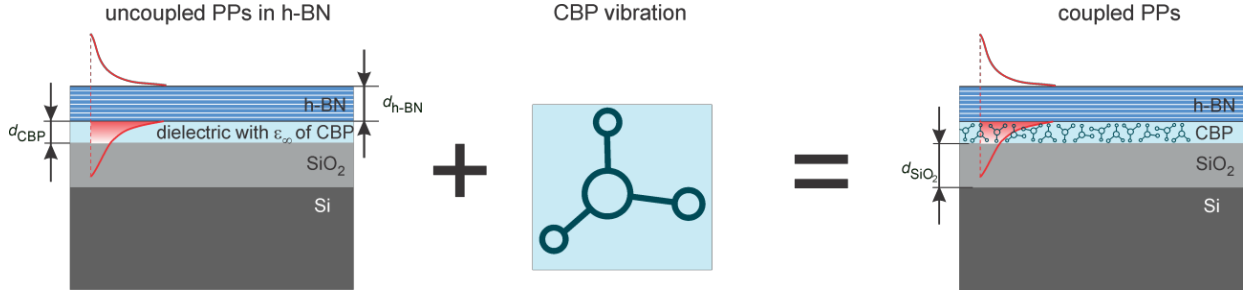

**Figure S8**. **Illustration of coupling between PPs and molecular vibrations.** (left) Illustration of the PP mode in a h-BN flake on top of Si/SiO$_2$ substrate, (center) illustration of the CBP molecule, (right) illustration of the coupled system involving PP mode and molecular vibration in the h-BN/CBP layers on top of Si/SiO$_2$ substrate.

We calculate eigenmodes (poles of $r_\mathrm{p}(q, \omega\text{-}i\gamma)$) considering the complex frequencies and real momenta of uncoupled and coupled PPs using the transfer matrix approach (see Methods). Eigenmode analysis considering complex frequencies automatically reveal half of the spectral linewidth of mode $\Gamma/2$, which is equal to the imaginary part of the eigenfrequency (temporal decaying $\gamma$) $\Gamma/2 = \gamma$.

Eigenmode analysis with complex frequencies of the coupled system yields the quasi-normal modes and splitting of the PP dispersion into two branches. We extract the mode splitting $\Omega$ as the smallest vertical separation between the real frequencies of the branches, which appears at the momentum corresponding to the intersection of dispersion considering complex momenta and the frequency of molecular vibration (see thin dashed black lines Fig. 3b,e of the main text).

Finally, we use strict mathematical condition to determine the strong coupling between PPs and molecular vibration[14]:

$$C \stackrel{\text{def}}{=} \frac{\Omega^2}{\frac{\Gamma_{\mathrm{CBP}}^2}{2} + \frac{\Gamma_{\mathrm{PP}}^2}{2}} > 1 \qquad (S7)$$

Note that for simplicity we calculate uncoupled PP linewidth, $\Gamma_{PP}$, at $\omega$ = 1450 cm$^{-1}$ and assume that the linewidth of the uncoupled PPs is the same in the considered range ($\omega_{PP} \in$ [1440, 1460] cm$^{-1}$).

**Strong coupling analysis between PP and CBP vibration in Fig. 2 of the main text**

We perform theoretical analysis of the coupling between PP and molecular vibration in h-BN/CBP layers in top of the Si/SiO$_2$ substrate (85 nm-thick h-BN with 100 nm-thick CBP layer below, SiO$_2$ thickness is 150 nm). The calculated quasi-normal modes considering the complex frequencies reveals the mode splitting $\Omega$ = 10 cm$^{-1}$ (white lines, Fig. 2e), linewidth of the uncoupled PP $\Gamma_{PP}$ = 8 cm$^{-1}$. Linewidth of the CBP vibration mode at 1450 cm$^{-1}$, $\Gamma_{CBP}$ = 6.4 cm$^{-1}$. With these parameters, mathematical condition of the strong coupling $C \stackrel{\text{def}}{=} \frac{\Omega^2}{\frac{\Gamma_{CBP}^2}{2}+\frac{\Gamma_{PP}^2}{2}} = \frac{10^2}{\frac{6.4^2}{2}+\frac{8^2}{2}} = 1.9 > 1$ are fulfilled.

Our numerical calculation demonstrates slightly smaller mode splitting ($\Omega$ = 10 cm$^{-1}$) than the analysis of experimental data ($\Omega$ = 11 cm$^{-1}$, see Supplementary information SIV). Potential reason for that could be the uncertainties in the h-BN, CBP and substrate dielectric functions. Particularly, the oscillator parameter $\omega_{p,1}$ = 128 cm$^{-1}$ that we determined for the CBP vibration at $\omega_{0,1}$ = 1450 cm$^{-1}$ (see Table S1) is slightly underestimated compared to literature values[9].

**Strong coupling analysis between PP and CBP vibration in Fig. 3 of the main text**

We perform theoretical analysis of the coupling between PPs in 1, 5 and 35 nm h-BN flake and molecular vibration in CBP layers of various thicknesses which lies on top of the Si/SiO$_2$ substrate (SiO$_2$ thickness is 250 nm). We calculated quasi-normal mode of the coupled PPs in h-BN/CBP layers and extracted the mode splitting, $\Omega$, as a smallest vertical separation between real frequencies of the low and upper branches.

Dielectric function of SiO$_2$ (see methods) reveals the imaginary part, which reduces the lifetime of the PPs. For our analysis in Fig. 3g we fix the linewidth of uncoupled PPs, $\Gamma_{pp}$, in the worst scenario when the h-BN flake lies on top of the SiO$_2$. We find that the linewidth of PPs in h-BN does not change much with thickness of h-BN flakes, see Table S3.

We estimate the minimal mode splitting, $\Omega_{min}$, for achieving the strong coupling using the mathematical condition Eq. S7, when C > 1. Minimal mode splitting for the strong coupling between PPs in the h-BN slab of different thicknesses and CBP molecular vibration at 1450 cm$^{-1}$ are reported in the Table S3.

| h-BN thickness [nm] | $\Gamma_{PP}$ ($d_{CBP} = 0$) [cm$^{-1}$] | $\Omega_{min}$ [cm$^{-1}$] |
|---|---|---|
| 35 | 2.6 | 5 |
| 5 | 2.5 | 4.9 |
| 1 | 2.5 | 4.9 |

**Table S3**. Linewidth of uncoupled PPs in h-BN of various thickness and minimal splitting for the strong-coupling between PPs and CBP vibration at 1450 cm$^{-1}$.